\documentclass[graybox]{svmult}
\usepackage{amscd}

\usepackage{amsmath}
\usepackage{amssymb}

\newtheorem{lem}{Lemma}
\newtheorem{pro}{Proposition}
\newtheorem{cor}{Corollary}
\newtheorem{exe}{Example}
\newtheorem{defi}{Definition}

\newtheorem{examp}{Example}

\def\sp{\hspace{0.2cm}}
\def\qed{\begin{flushright} $\square$ \end{flushright}}
\def\qee{\begin{flushright} $\Diamond$ \end{flushright}}

\begin{document}

\title*{A dynamical point of view of Quantum Information: Wigner measures}
\titlerunning{A dynamical point of view of Quantum Information}

\author{A. Baraviera, C. F. Lardizabal,  A. O. Lopes, and M. Terra Cunha}

\institute{A. T. Baraviera \at I.M. - UFRGS, Porto Alegre - 91500-000, Brasil, \email{atbaraviera@gmail.com}
\and C. F. Lardizabal \at I.M. - UFRGS, Porto Alegre - 91500-000, Brasil, \email{carlos.lardizabal@gmail.com }
\and A. O. Lopes \at I.M. - UFRGS, Porto Alegre - 91500-000, Brasil, \email{arturoscar.lopes@gmail.com}
\and M. Terra Cunha  \at D. M - UFMG, Belo Horizonte - 30161-970, Brasil, \email{marcelo.terra.cunha@gmail.com}
}

\maketitle

\abstract{We analyze a known version of the discrete Wigner function and some connections with Quantum Iterated Funcion Systems.}

\bigskip
Dynamics, Games and Science II, DYNA 2008, Edit. M. Peixoto, A. Pinto and D Rand, Springer Verlag (2011)

\section{Discrete Weyl relations}\label{sec_rel_weyl}

This section follows parts of \cite{benatti}. Consider the Hilbert space $\mathcal{H}=\mathbb{C}^N$. Let $\{\vert k\rangle\}_{k=0}^{N-1}$ be an orthonormal base. Fix $\alpha_{u},\alpha_{v}\in [0,1]$ and define the following matrices $U_N$, $V_N\in M_N(\mathbb{C})$:
\begin{equation}\label{pos_mom_displacement}
U_N:=e^{\frac{2\pi}{N}i\alpha_u}\sum_{k=0}^{N-1}e^{\frac{2\pi}{N}ik}\vert k\rangle\langle k\vert, \sp V_N:=e^{\frac{2\pi}{N}i\alpha_v}\sum_{k=0}^{N-1}\vert k\rangle\langle k-1\vert
\end{equation}
together with the identification $\vert j\rangle=\vert j \textrm{ mod } N\rangle$. Such operators are unitary and we have
\begin{equation}\label{u,v_prop}
U_N\vert l\rangle=e^{\frac{2\pi}{N}i(\alpha_u+l)}\vert l\rangle,\sp V_N\vert l\rangle=e^{\frac{2\pi}{N}i\alpha_v}\vert l+1\rangle
\end{equation}
Defining $n:=(n_1,n_2)\in\mathbb{Z}^2$, we have that $U_N$ and $V_N$ satisfy the {\bf discrete Weyl relations}
\begin{equation}\label{weyl_discretas}
U_N^{n_1}V_N^{n_2}=e^{\frac{2\pi}{N}in_1n_2}V_N^{n_1}U_N^{n_2}
\end{equation}
Also, inspired in the continuous case, we define the {\bf discrete Weyl operators}:
\begin{equation}\label{weyl_disc}
W_N(n):=e^{-i\frac{\pi}{N}n_1n_2}U_N^{n_1}V_N^{n_2}
\end{equation}
Such operators satisfy
\begin{equation}
W_N^* (n)=W(-n)
\end{equation}
and
\begin{equation}
W_N(n)W_N(m)=e^{i\frac{\pi}{N}\sigma(n,m)}W_N(n+m)
\end{equation}
where $\sigma(n,m):=n_1m_2-n_2m_1$.

\bigskip

When normalized, the discrete Weyl operators form an orthonormal base for $M_N(\mathbb{C})$. In fact, using (\ref{u,v_prop}) and (\ref{weyl_disc}), we have
$$tr(W_N(n))=\sum_{l=0}^{N-1}e^{-i\frac{\pi}{N}n_1n_2}\langle l\vert U_N^{n_1}V_N^{n_2}\vert l\rangle$$
$$=\sum_{l=0}^{N-1}e^{-i\frac{\pi}{N}(n_1n_2+2n_1(\alpha_u+l)-2n_2\alpha_v)}\langle l\vert l+n_2\rangle$$
\begin{equation}
=\delta_{n_2,0}\sum_{l=0}^{N-1}e^{-\frac{2\pi i n_1}{N}(\alpha_u+l)}=N\delta_{n,0}
\end{equation}
This allows us to obtain
\begin{equation}
tr(W_N^*(n)W_N(m))=N\delta_{n,m}
\end{equation}
and therefore for all $A\in M_N(\mathbb{C})$,
\begin{equation}
A=\frac{1}{N}\sum_{n\in\mathbb{Z}_N^2} tr\Big(W_N^*(n)A\Big)W_N(n)
\end{equation}
where $\mathbb{Z}_N^2:=\{n=(n_1,n_2):0 \leq n_i\leq N-1\}$.

\qee

\section{Introduction to the Wigner function}

This section follows parts of \cite{tannor}. Given a quantum system, we are interested in obtaining another form of representing the wave function $\Psi(x)$. Such object will be the Wigner function, which will depend on two variables, moment and position. In order to understand such functions, we need to study the structure of phase spaces.

\bigskip

The Wigner function consists of a special way of describing density operators. In principle, we could say that density operators are a more fundamental structure than its Wigner representation. For instance, the Wigner representation is unable to describe the density operators associated to two-level systems. However, due to its simplicity, we will see that an understanding of the Wigner distribution gives us insight on certain aspects of density operators.

\bigskip

\begin{defi}
Given a wave function $\Psi(x)$, the {\bf Wigner distribution function} is
\begin{equation}
W(q,p)=W_\Psi(q,p)
:=\frac{1}{2\pi\hbar}\int_{-\infty}^{\infty}e^{isp/\hbar}\langle
q-\frac{s}{2}\vert\Psi\rangle\langle\Psi\vert q+\frac{s}{2}\rangle
ds
\end{equation}
where above we are using Dirac notation
\begin{equation}
\langle q-\frac{s}{2}\vert\Psi\rangle=\Psi(q-\frac{s}{2})
\end{equation}
\begin{equation}
\langle\Psi\vert q+\frac{s}{2}\rangle=\Psi^*(q+\frac{s}{2})
\end{equation}
\end{defi}
Define the change of coordinates
\begin{equation}
x=q+\frac{s}{2},\sp x'=q-\frac{s}{2}
\end{equation}
and then we obtain
\begin{equation}\label{wigner_newcoord}
W(q,p)=\frac{1}{2\pi\hbar}\int_{-\infty}^{\infty}e^{\frac{i}{\hbar}p(x-x')}\langle x'\vert\Psi\rangle\langle\Psi\vert x\rangle ds
\end{equation}
That is, the Wigner distribution is obtained by calculating the product
$\Psi(x')\Psi^*(x)$ and then applying the Fourier transform on $s=x-x'$. Such distribution has the following properties:
\begin{equation}
\int_{-\infty}^{\infty}W(q,p)dp=\langle q\vert\Psi\rangle\langle\Psi\vert q\rangle=\vert \Psi(q)\vert^2
\end{equation}
\begin{equation}
\int_{-\infty}^{\infty}W(q,p)dq=\langle p\vert\Psi\rangle\langle\Psi\vert p\rangle=\vert \tilde{\Psi}(p)\vert^2
\end{equation}
\begin{equation}
\int_{-\infty}^{\infty}\int_{-\infty}^{\infty}W(q,p)dpdq=1
\end{equation}
where $\tilde{\Psi}$ is the moment representation of the wave function $\Psi$.

\bigskip

The Wigner function is real, but can assume negative or positive values. In this sense, it is not a density, but it is a kind of joint distribution of the position and momentum
distributions.

Now, note that (\ref{wigner_newcoord}) can be written as
\begin{equation}\label{wigner_basic_not}
W(q,p)=\frac{1}{2\pi\hbar}\int_{-\infty}^{\infty}e^{\frac{i}{\hbar}p(x-x')}\langle x'\big(\vert\Psi\rangle\langle\Psi\vert \big)x\rangle ds=\frac{1}{2\pi\hbar}\int_{-\infty}^{\infty}e^{\frac{i}{\hbar}p(x-x')}\langle x'\vert\rho\vert x\rangle ds
\end{equation}
where
\begin{equation}
x=q+\frac{s}{2},\sp x'=q-\frac{s}{2}
\end{equation}
where we define the density operator associated to a pure state as
\begin{equation}
\rho:=\vert\Psi\rangle\langle\Psi\vert
\end{equation}
The general definition for  $\rho$ includes pure and mixed states:
\begin{equation}\label{dbd1_1}
\rho=\sum_i p_i\vert\Psi_i\rangle\langle \Psi_i\vert
\end{equation}
where $p_i\geq 0$ and $\sum_i p_i=1$. Such equation describes $\rho$ as an incoherent  superposition of pure state density operators $\vert\Psi_i\rangle\langle\Psi_i\vert$, where $\Psi_i$ is a wave function, but not necessarily an energy eigenstate. On equation (\ref{dbd1_1}) the $p_i$ denote the probabilities of finding the system on the state $\vert\Psi_i\rangle$.

\bigskip

Hence, besides the usual probabilistic interpretation for finding a particle described by a certain wave function at some position, we also have a probability distribution that such a particle can be found in different states.

\qee

\section{Discrete Wigner function}\label{wig_discreta}

This section follows parts of \cite{miquel} and \cite{wootters}. In dimension 1, the {\bf continuous Wigner function} is in 1-1 correspondence with a density matrix $\rho$ and is defined by
\begin{equation}
W_\rho(q,p)=W(q,p):=\frac{1}{2\pi\hbar}\int_{-\infty}^{\infty}
e^{i\lambda p/\hbar}\langle q-\frac{\lambda}{2}\vert\rho\vert
q+\frac{\lambda}{2}\rangle d\lambda
\end{equation}
Such function is uniquely defined by the following properties:
\cite{miquel},\cite{wootters}:
\begin{enumerate}
\item $W(q,p)\in\mathbb{R}$
\item If $\rho_1$ and $\rho_2$ are two density states then
\begin{equation}
tr(\rho_1\rho_2)=2\pi\hbar\int W_1(q,p)W_2(q,p)dqdp
\end{equation}
\item (Projection property) The integral along a line on phase space, described by $a_1q+a_2p=a_3$, is the probability density that the measurement of the observable $a_1\hat{Q}+a_2\hat{P}$ gives $a_3$ as a result.
\end{enumerate}

\bigskip

{\bf Remark} Note that the Wigner function is always associated to
a density matrix. It would be more appropriate to use the notation
$W_\rho$ instead of $W$. When there is no possibility of confusion
we will denote $W$. The projection property stated above means, in
other words, that the projection of the Wigner function along any
direction of the phase space is equal to the probability
distribution of a certain observable $a_1q+a_2p$, associated to
that direction. Two special cases of this property are well-known:
\begin{equation}
\int W(q,p)dq
\end{equation}
is the probability distribution for the moment, and
\begin{equation}
\int W(q,p)dp
\end{equation}
is the probability distribution for position. For more details on these properties, see \cite{wootters}.
\qee

We can write $W$ as the expected value of a Fano operator, so we have
\begin{equation}\label{def_wigner_1}
W(q,p)=tr(\rho \hat{A}(q,p))
\end{equation}
where $\hat{A}$ can be written as
\begin{equation}
\hat{A}(q,p)=\frac{1}{(2\pi\hbar)^2}\int exp\Big[-\frac{\lambda}{\hbar}(\hat{P}-p)+i\frac{\lambda'}{\hbar}(\hat{Q}-q)\Big]d\lambda d\lambda'
\end{equation}
\begin{equation}
=\frac{1}{(2\pi\hbar)^2}\int \hat{D}(\lambda,\lambda')exp\Big[-\frac{i}{\hbar}(\lambda' q-\lambda p)\Big]d\lambda d\lambda'
\end{equation}
where
\begin{equation}
\hat{D}(\lambda,\lambda'):=exp\Big[-\frac{i}{\hbar}(\lambda\hat{P}-\lambda'\hat{Q})\Big]
\end{equation}
Also we can write $\hat{A}$ as
\begin{equation}
\hat{A}(q,p)=\frac{1}{\pi\hbar}\hat{D}\hat{R}\hat{D}^*
\end{equation}
where above we write $\hat{D}=\hat{D}(q,p)$ and $\hat{R}$ is an operator acting on positive eigenstates such that $\hat{R}\vert x\rangle=\vert-x\rangle$.

\bigskip

The proof that $W$ satisfies properties 1 to 3 stated above follows from simple phase space properties. The fact that $W(q,p)\in\mathbb{R}$ is a consequence of the fact that  $\hat{A}(q,p)$ is hermitian.  As for property 2, we can show that
\begin{equation}
tr\Big(\hat{A}(q,p)\hat{A}(q',p')\Big)=\frac{1}{2\pi\hbar}\delta(q-q')\delta(p-p')
\end{equation}
As a consequence, it is possible to invert equation (\ref{def_wigner_1}) so we can write
\begin{equation}
\rho=2\pi\hbar\int W(q,p)\hat{A}(q,p)dq dp
\end{equation}
Property 2 follows from the formula above. As for property 3, note that by integrating $\hat{A}(q,p)$ along a line on phase space gives us a projection operator. Therefore
\begin{equation}
\int \delta(a_1q+a_2p-a_3)\hat{A}(q,p) dq dp=\vert a_3\rangle\langle a_3\vert
\end{equation}
where $\vert a_3\rangle$ is an eigenstate of the operator $a_1\hat{Q}+a_2\hat{P}$ with eigenvalue $a_3$. Later we will describe the proof of this property for the discrete case.

\qee

Now we are interested in defining the Wigner function in the discrete case. The first step is to define a discrete phase space. Consider a Hilbert space of dimension $N$ and define a base
$$B_x=\{\vert n\rangle, n=0,\dots, N-1\},$$
which will be seen as a {\bf discrete position base}. Now we define a base of moments
$$B_p=\{\vert k\rangle, k=0,\dots, N-1\}$$
A natural way of introducing the moment base from the position
base is via the {\bf discrete Fourier transform}. Then we can
obtain the states of $B_p$ from the states in $B_x$ in the
following way:
\begin{equation}
\vert k\rangle=\frac{1}{\sqrt{N}}\sum_n \exp[ 2\pi i nk/N] \vert n\rangle
\end{equation}
Therefore, as in the continuous case, position and moment are related by the Fourier transform.

\bigskip
{\bf Remark} We can relate the dimension of the Hilbert space with the Planck constant in the following way. We are supposing that the phase space has a finite area, which we can suppose equal to 1. In this area we can have $N$ orthogonal states. If each state fills an area equal to $2\pi\hbar$, we have $N=1/2\pi\hbar$. So $N$ plays the role of the inverse of the Planck constant and the limit as $N$ goes to infinity can be seen as the semiclassical limit \cite{miquel}.

\qee

Given position and moment bases, we can define their respective displacement operators.
For discrete systems, we can define translation operators $\hat{U}$ and $\hat{V}$, in a way which is similar to what we have in (\ref{pos_mom_displacement}) and (\ref{u,v_prop}), section \ref{sec_rel_weyl}:
\begin{equation}\label{cwig_1}
\hat{U}^m\vert n\rangle:=\vert n+m\rangle,\sp \hat{U}^m\vert k\rangle:=\exp[-2\pi imk/N]\vert k\rangle
\end{equation}
where the vector sums are $\textrm{mod } N$. In a similar way the operator  $\hat{V}$ is a shift on moment base, and it is diagonal on positions:
\begin{equation}\label{cwig_2}
\hat{V}^m\vert k\rangle:=\vert k+m\rangle,\sp \hat{V}^m\vert n\rangle:=\exp[2\pi imn/N]\vert n\rangle
\end{equation}
Then it is possible to show that
\begin{equation}
\hat{V}^p\hat{U}^q=e^{2\frac{\pi}{N}ipq}\hat{U}^q\hat{V}^p,
\end{equation}
the discrete Weyl relations (\ref{weyl_discretas}), seen on section \ref{sec_rel_weyl}. Let us also define a reflection operator  as $\hat{R}\vert n\rangle:=\vert -n\rangle$. We have that
\begin{equation}
\hat{U}\hat{R}=\hat{R}\hat{U}^{-1}, \sp \hat{V}\hat{R}=\hat{R}\hat{V}^{-1}
\end{equation}
The reflection operator is related to the Fourier transform in the following way.
Denote by $U_{FT}$ the discrete Fourier transform, that is the operator whose entries on base $B_x$ are
\begin{equation}
\langle n'\vert U_{FT}\vert n\rangle=\exp[2\pi inn'/N]
\end{equation}
Then we have
\begin{equation}
\hat{R}=U_{FT}^2
\end{equation}

\qee

In order to define the discrete Wigner function, we still have to define a translation operator $\hat{T}$ and a point operator $\hat{A}$, corresponding to the Fano operator defined in the continuous case. This is what we will do next. Define
\begin{equation}
\hat{T}(q,p):=\hat{U}^q\hat{V}^p\exp[i\pi qp/N]
\end{equation}
Such operators satisfy
\begin{equation}
\hat{T}(\lambda q,\lambda p)=\hat{T}^{\lambda}(q,p)
\end{equation}

{\bf Remark} In $\mathbb{R}^2$ we define the translation operator with position $q$ and moment  $p$ as
\begin{equation}\label{em_r2_trans}
\hat{T}(q,p)=e^{-\frac{i}{\hbar}(q\hat{P}-p\hat{Q})}
\end{equation}
Instead of definitions (\ref{cwig_1}) and (\ref{cwig_2}) given for $\hat{U}$ and $\hat{V}$ we could, in principle, define $\hat{U}$ and $\hat{V}$ as the exponential of two operators $\hat{Q}$ and $\hat{P}$, defined as being diagonal in $B_x$ and $B_p$. However, infinitesimal operators $\hat{Q}$ and $\hat{P}$ satisfying the canonical commutation relations (CCR) cannot be defined over a discrete Hilbert space \cite{garciamata},\cite{wootters}. Because of that we will use the finite cyclic shifts, given by (\ref{cwig_1}) and (\ref{cwig_2}).

\qee

{\bf Remark} Due to technicalities, the phase-space can be taken to be a $N\times N$ or a $2N\times 2N$ grid \cite{miquel}. Typically we will be interested in phase spaces with even dimension and we will use the $2N\times 2N$ grid (for instance, if $N=2$ the phase space has 16 points). Our following definitions will follow this choice as well.

\qee

Let $\alpha=(q,p)$ be a point of the discrete phase space, with $q$ and $p$ assuming values between 0 and $2N-1$. Define
\begin{equation}\label{expr_para_hat_a}
\hat{A}(\alpha):=\frac{1}{(2N)^2}\sum_{\lambda,\lambda'=0}^{2N-1}\hat{T}(\lambda,\lambda')\exp\Big[-2\pi i\frac{(\lambda'q-\lambda p)}{2N}\Big]=\frac{1}{2N}\hat{U}^q\hat{R}\hat{V}^{-p}e^{i\pi pq/N}
\end{equation}
We can express the translation operator in terms of $\hat{A}(\alpha)$ by inverting the above definition and then we obtain the Fourier transform of $\hat{A}$:
\begin{equation}
\tilde{T}(n,k)=\sum_{q,p=0}^{2N-1}\hat{A}(q,p)\exp[-i\frac{2\pi}{2N}(np-kq)]
\end{equation}
Note that as we defined the point operators over a lattice of $2N\times 2N$ points, we get a total of $4N^2$ operators. However, such set is not independent. That is, we can show that
\begin{equation}
\hat{A}(q+\sigma_q N,p+\sigma_p N)=\hat{A}(q,p)(-1)^{\sigma_p q+\sigma_q p+\sigma_q\sigma_p N}
\end{equation}
for $\sigma_q,\sigma_p=0,1$. So we have that $N^2$ operators define the remaining ones.
Define
$$G_N:=\{\alpha=(q,p):0\leq q,p\leq N-1\}$$
And the set  $G_{2N}$ will denote the entire lattice of order $2N$.

\bigskip

A relation between $\hat{A}$ and $\hat{T}$ is the following:
\begin{equation}
\hat{A}(\alpha)\hat{A}(\alpha')=\hat{T}(\alpha-\alpha')\frac{\exp[i(\pi/N)(q_\alpha p_{\alpha'}-q_{\alpha'}p_{\alpha})]}{4N^2}
\end{equation}
By taking the trace of the above equation we get
\begin{equation}
tr(\hat{A}(\alpha)\hat{A}(\alpha'))=\frac{1}{4N}\delta_N(q'-q)\delta_N(p'-p)
\end{equation}
where $\alpha$ and $\alpha'$ are in $G_N$ and

\begin{equation}
\delta_N(q):=\frac{1}{N}\sum_{n=0}^{N-1}e^{-2\pi iqn/N}
\end{equation}
is the periodic Dirac delta function, which is equal to zero unless $q\equiv 0\textrm{ mod }N$.

\bigskip

\begin{defi}
The {\bf discrete Wigner function} is
\begin{equation}
W(\alpha)=W_\rho (\alpha):=tr(\hat{A}(\alpha)\rho)
\end{equation}
where $\alpha\in G_{2N}$.
\end{defi}
These $4N^2$ values are not independent because in a similar way to what we have for the operator $\hat{A}$, we have
\begin{equation}\label{reconst_d_wigner}
\hat{W}(q+\sigma_q N,p+\sigma_p N)=\hat{W}(q,p)(-1)^{\sigma_p q+\sigma_q p+\sigma_q\sigma_p N}
\end{equation}
for $\sigma_q,\sigma_p=0,1$. As the operators $\hat{A}(\alpha)$ form a complete set, we can write the density operator as a linear combination of the $\hat{A}(\alpha)$. So we can show that
\begin{equation}\label{exp_para_rho_wigner}
\rho=4N\sum_{\alpha\in G_N} W(\alpha)\hat{A}(\alpha)=N\sum_{\tilde{\alpha}\in G_{2N}} W(\tilde{\alpha})\hat{A}(\tilde{\alpha})
\end{equation}

{\bf Remark} It is possible to show that the discrete Wigner function defined above satisfies properties 1 to 3 stated in the beginning of this section. Property 1 is a consequence of the fact that $\hat{A}(q,p)$ are hermitian operators. Property 2 follows from the completeness of the set $\hat{A}(\alpha)$, which allows us to show that
\begin{equation}
tr(\rho_1\rho_2)=N\sum_{\alpha\in G_{2N}}W_1(\alpha)W_2(\alpha)
\end{equation}
The proof of the third property requires a brief analysis of the lattice $G_N$ and we refer the reader to section 21 for details.

\qee

{\bf Conclusions} We have defined the Wigner function for systems over a Hilbert space of dimension $N<\infty$. The Wigner functions is defined as the expected value of the operator $\hat{A}(\alpha)$ defined over the phase space given by equation (\ref{expr_para_hat_a}). The definition is such that $W(\alpha)\in\mathbb{R}$ and is such that we can calculate the inner product between states and gives the correct marginal distributions along any line over the phase space, which is the lattice $G_{2N}$ with $4N^2$ points. Also, the values of $W(\alpha)$ on the sublattice $G_N$ are enough to determine $W$ in the entire space.

\qee

\section{Calculating Wigner functions}\label{calc_f_w_sec}

In order to calculate the Wigner function of a quantum state, we will use
(\ref{cwig_1}),  (\ref{cwig_2}) and (\ref{expr_para_hat_a}) se we can write $W$ in the following convenient form:

\begin{lem}
\begin{equation}
W(q,p)=\frac{1}{2N}\sum_{n=0}^{N-1}\langle q-n\vert\rho\vert n\rangle\exp\Big[\frac{2\pi i}{N}p(n-q/2) \Big]
\end{equation}
\end{lem}
{\bf Proof}  In the following calculations, recall that the inner
product is linear on the second variable. We have that
$$W(q,p)=tr(A\rho)=\frac{1}{2N}\exp[i\pi pq/N]tr(U^qRV^{-p}\rho)$$
$$=\frac{1}{2N}\exp[i\pi pq/N]\sum_{i=0}^{N-1}\langle n\vert U^qRV^{-p}\rho \vert n\rangle=\frac{1}{2N}\exp[i\pi pq/N]\sum_{i=0}^{N-1}\langle U^{-q} n\vert RV^{-p}\rho\vert n\rangle$$
$$=\frac{1}{2N}\exp[i\pi pq/N]\sum_{i=0}^{N-1}\langle n-q\vert RV^{-p}\rho\vert n\rangle=\frac{1}{2N}\exp[i\pi pq/N]\sum_{i=0}^{N-1}\langle q-n\vert V^{-p}\rho\vert n\rangle$$
$$=\frac{1}{2N}\exp[i\pi pq/N]\sum_{i=0}^{N-1}\langle V^p(q-n)\vert \rho\vert n\rangle=$$
$$
\frac{1}{2N}\exp[i\pi pq/N]\sum_{i=0}^{N-1} \exp[-2\pi ip(q-n)/N]\langle q-n\vert \rho\vert n\rangle$$
Also, note that
$$i\pi pq/N -2\pi ip(q-n)/N=\frac{ip\pi}{N}(q-2(q-n))=\frac{ip\pi}{N}(2n-q)=\frac{2\pi ip}{N}(n-q/2)$$
Hence,
$$W(q,p)=\frac{1}{2N}\sum_{n=0}^{N-1} \langle q-n\vert \rho\vert n\rangle\exp[\frac{2\pi ip}{N}(n-q/2)]$$

\qed

We believe there is a misprint in  \cite{miquel} in the expression corresponding to the $W(q,p)$ described by the claim of the above lemma.

One can ask how $W_\rho$ changes with $\rho$. Suppose first $\rho$ is a projector from a wave $\psi$ which has norm 1 in ${\cal L}^2$. Suppose $ (a \phi_1 + b \phi_2) = \psi$,
where $\psi, \phi_1, \phi_2$ have norm $1$, and $\rho =
|\psi><\psi|$. Then, $W_\psi\neq a W_{\phi_1} + b W_{\phi_1}.$ The linearity occurs only when  $\rho = \sum_i c_i | i >< i|$,
that is, when $\rho$ is diagonal. This in general do not happen
for operators $|\psi><\psi|$ induced by a wave $\psi$. However, if
$\rho= (a \rho_1 + b \rho_2)  $, where  $\rho, \rho_1, \rho_2$ are
density matrices, then $ W_\rho = a W_{\rho_1} + b W_{\rho_2}$.

\begin{exe}\label{exe_int_wigner}
Let $N=2$, and let $\vert\psi\rangle=a\vert 0\rangle+b\vert
1\rangle$ be a state superposition. Let $W_1(\alpha)$ and
$W_2(\alpha)$ be the Wigner functions for $\vert 0\rangle$ and
$\vert 1\rangle$, respectively. We have that the Wigner function $W$ for $\vert\psi\rangle$ is such that
\begin{equation}
W(\alpha)=\vert a\vert^2 W_1(\alpha)+\vert b\vert^2
W_2(\alpha)+2Re\{ab^*\langle 1\vert A(\alpha)\vert 0\rangle\}
\end{equation}
In fact, note that
$$W(\alpha)=tr(A(\alpha)\rho)=tr\Big(A(\alpha)(\vert a\vert^2\vert 0\rangle\langle 0\vert+\vert b\vert^2\vert 1\rangle\langle
1\vert+ab^*\vert 0\rangle\langle 1\vert+a^*b\vert 1\rangle\langle
0\vert) \Big)$$
$$=\vert a\vert^2 W_1(\alpha)+\vert
b\vert^2W_2(\alpha)+ab^*tr(A(\alpha)\vert 0\rangle\langle
1\vert)+a^*btr(A(\alpha)\vert 1\rangle\langle 0\vert)$$
$$=\vert a\vert^2 W_1(\alpha)+\vert
b\vert^2W_2(\alpha)+ab^*tr(\langle 1\vert A(\alpha)\vert
0\rangle)+a^*b tr(\langle 0\vert A(\alpha)\vert 1\rangle)$$
so the result follows.
\end{exe}
\qee

Let us remark a few properties of the Wigner function for a pure state $\rho$. In this case by expanding $\rho$ in terms of the phase space operators as in equation (\ref{exp_para_rho_wigner}) and by imposing the condition $\rho^2=\rho$, we get
\begin{equation}\label{restri_evol}
W(\alpha)=4N^2\sum_{\beta,\gamma\in G_{N}}\Gamma(\alpha,\beta,\gamma)W(\beta)W(\gamma)
\end{equation}
where the function $\Gamma(\alpha,\beta,\gamma)$, which depends on 3 points (i.e., a triangle) is given by
\begin{equation}
\Gamma(\alpha,\beta,\gamma):=tr(\hat{A}(\alpha)\hat{A}(\beta)\hat{A}(\gamma))=\frac{1}{4N^3}\exp\Big[ \frac{2\pi i}{N}S(\alpha,\beta,\gamma) \Big],
\end{equation}
of 2 of the 3 point $(\alpha,\beta,\gamma)$ contain even $q$ and $p$ coordinates. Otherwise we define
\begin{equation}
\Gamma(\alpha,\beta,\gamma):=0,
\end{equation}
and in the above expression, valid for even $N$, the value $S(\alpha,\beta,\gamma)$ is the area of the triangle formed by these points (measured in units of the elementary triangle formed by 3 points which are one position apart from each other).

\qee

Now we calculate the Wigner function for a position eigenvalue
\begin{equation}
\rho_{q_0}=\vert q_0\rangle\langle q_0\vert
\end{equation}
We obtain the following closed expression for $W$:
$$W_{q_0}(q,p)=\frac{1}{2N}\langle q_0\vert \hat{U}^q\hat{R}\hat{V}^{-p}\vert q_0\rangle e^{i\pi pq/N}$$
\begin{equation}\label{fech_puros}
=\frac{1}{2N}\delta_N(q-2q_0)(-1)^{p[(q-2q_0) \textrm{ mod } N]}
\end{equation}
We can also write the Wigner function of a state which is a linear superposition:
\begin{equation}
\vert\psi\rangle=\frac{1}{\sqrt{2}}(\vert q_0\rangle+e^{-i\phi}\vert q_1\rangle)
\end{equation}
Again, we can obtain a closed expression for $W$, which is
\begin{equation}\label{fech_mist}
W(q,p)=\frac{1}{2}\Big(W_{q_0}(q,p)+W_{q_1}(q,p)+\Delta W_{q_0,q_1}(q,p)\Big)
\end{equation}
where the interference term is
\begin{equation}
\Delta W_{q_0,q_1}(q,p):=\frac{1}{N}\delta_N(\tilde{q})(-1)^{\tilde{q}p}\cos\Big(\frac{2\pi}{\lambda}p+\phi \Big)
\end{equation}
where
\begin{equation}
\tilde{q}=q_0+q_1-q, \sp \lambda=\frac{2N}{q_0-q_1}
\end{equation}
This is an explicit expression for the calculation seen in example \ref{exe_int_wigner}.
\qee

Now we make a few considerations on the time evolution of quantum systems on phase space.
If $U$ is the unitary operator which determines the evolution of a state, then the associated density matrix evolves in the following way,
\begin{equation}
\rho(t+1)=U\rho(t) U^*
\end{equation}
By this fact, we can show that the Wigner function evolves in the following way:
\begin{equation}
W(\alpha,t+1)=\sum_{\beta\in G_{2N}} Z_{\alpha\beta}W(\beta, t)
\end{equation}
where the matrix $Z_{\alpha\beta}$ is defined as
\begin{equation}
Z_{\alpha\beta}:=N tr\Big(\hat{A}(\alpha)U\hat{A}(\beta)U^*\Big)
\end{equation}
Therefore the time evolution in phase space is represented by a linear transformation, which is a consequence of Schr\"odinger's equation. The unitarity imposes a few restrictions on the matrix $Z_{\alpha\beta}$. In fact, since purity of states is preserved, the time evolution has to preserve the restriction given by equation (\ref{restri_evol}). Therefore, the matrix has to leave the function $\Gamma(\alpha,\beta,\gamma)$ invariant, that is,
\begin{equation}
\Gamma(\alpha',\beta',\gamma')=\sum_{\alpha,\beta,\gamma}Z_{\alpha'\alpha}Z_{\beta'\beta}Z_{\gamma'\gamma}\Gamma(\alpha,\beta,\gamma)
\end{equation}
The real matrix $Z_{\alpha\beta}$ contains all the information on the time evolution of the system. In general, such matrix relates a point $\alpha$ with several other points $\beta$.
So the evolution will be, in general, nonlocal, a unique property of quantum mechanics. In classical systems, the value of the classical distribution function $W(\alpha,t+1)$ is equal to the value $W(\beta,t)$ for some point $\beta$, which consists of a well defined function of $\alpha$ and $t$. However, we have in \cite{miquel} a few examples of unitary operators which generate a local dynamical evolution on the phase space.

\qee

To conclude this section, we calculate the Wigner function for a quantum channel $\Lambda$, as the ones considered for our analysis of QIFS. This is a straightforward calculation. Let
$V_i$ be linear operators, $i=1,\dots, k$ such that $\sum_i V_i^*V_i=I$. Then $\Lambda(\rho)=\sum_i V_i\rho V_i^*\in\mathcal{M}_N$. Hence,
$$W_{\Lambda(\rho)}(q,p)=\frac{1}{2N}\sum_{n=0}^{N-1}\langle q-n\vert\Lambda(\rho)\vert n\rangle\exp\Big[\frac{2\pi i}{N}p(n-q/2) \Big]$$
$$=\frac{1}{2N}\sum_{n=0}^{N-1}\sum_{i=1}^k\langle q-n\vert V_i\rho V_i^* \vert n\rangle\exp\Big[\frac{2\pi i}{N}p(n-q/2) \Big]$$
\begin{equation}
=\frac{1}{2N}\sum_{n=0}^{N-1}\sum_{i=1}^k\langle (q-n) V_i\vert\rho\vert V_i^*  (n)\rangle\exp\Big[\frac{2\pi i}{N}p(n-q/2) \Big]
\end{equation}
Writing $\rho=\sum_{j=0}^{N-1}\rho_j\vert j\rangle\langle j\vert$, $\sum_j \rho_j=1$, we get
\begin{equation}
W_{\Lambda(\rho)}(q,p)=\frac{1}{2N}\sum_{n,j=0}^{N-1}\sum_{i=1}^k\rho_j\langle (q-n) V_i\vert j\rangle\langle j \vert V_i^*  (n)\rangle\exp\Big[\frac{2\pi i}{N}p(n-q/2) \Big]
\end{equation}
Therefore the Wigner function of $\Lambda(\rho)$ is obtained in a simple way from the function for $\rho$.

\qee

\section{some properties of the discrete Wigner function}

We have seen in section \ref{wig_discreta} that the discrete Wigner function
\begin{equation}
W(\alpha)=tr(\hat{A}(\alpha)\rho)
\end{equation}
satisfies properties 1 and 2. Now let us prove property 3. Let $\rho=\sum_i p_i\vert i\rangle\langle i\vert$, $\sum_i p_i=1$ be a density operator. Denote by
$$B_x=\{\vert n\rangle, n=0,\dots, N-1\},$$
a position basis and
$$B_p=\{\vert k\rangle, k=0,\dots, N-1\}$$
a moment basis, as before, where
\begin{equation}
\vert k\rangle=\frac{1}{\sqrt{N}}\sum_n \exp[ 2\pi i nk/N] \vert n\rangle
\end{equation}
To prove property 3, we must show that as we sum the operators $\hat{A}(q,p)$ over the point of the phase space which lie over a line $L$, we obtain a projection operator. This implies that by summing the values of the Wigner function over all the points of a line we get a positive number, which can be interpreted as a probability.

\bigskip

We begin by defining a line on the phase space. A line $L$ is a set of point of the lattice, defined as
\begin{equation}
L=L(n_1,n_2,n_3)=\{(q,p)\in G_{2N}: n_1p-n_2q=n_3, 0\leq n_i\leq 2N-1\}
\end{equation}
Also, we say that two lines as parallel if they are parameterized by the same integers $n_1$ and $n_2$.

\bigskip

Now, let us show that as we sum the point operators $A$ over a line, we get projection operators. So we are interested in the operator
\begin{equation}
\hat{A}_L=\sum_{(q,p)\in L} \hat{A}(q,p)
\end{equation}
Since $\delta_N(q)=\frac{1}{N}\sum_{n=0}^{N-1}e^{-2\pi iqn/N}$, we can rewrite such operator as $$A_L=\sum_{q,p=0}^{2N-1}\hat{A}(q,p)\delta_{2N}(n_1p-n_2q-n_3)$$
$$=\frac{1}{2N}\sum_{\lambda=0}^{2N-1}\sum_{q,p=0}^{2N-1}\hat{A}(q,p)\exp[-i\frac{2\pi}{2N}\lambda(n_1p-n_2q-n_3)]$$
\begin{equation}\label{al_exp1}
=\frac{1}{2N}\sum_{\lambda=0}^{2N-1}\hat{T}^{\lambda}(n_1,n_2)\exp[i\frac{2\pi}{2N}n_3\lambda]
\end{equation}
where we use the Fourier transform of $\hat{A}$ to obtain the last equality. Since $\hat{T}$ is unitary, we have $N$ eigenvectors $\vert \phi_j\rangle$ with eigenvalues $\exp[-2\pi i\phi_j/N]$. Besides, such operator is cyclic and satisfies $\hat{T}^N=I$. Therefore as its eigenvalues are $N-th$ roots of unity, the $\phi_j$ are integers. So we can rewrite (\ref{al_exp1}) as
$$\hat{A}_L=\frac{1}{2N}\sum_{\lambda=0}^{2N-1}\sum_{j=0}^N\exp[-i\frac{2\pi}{2N}(2\phi_j-n_3)\lambda]\vert\phi_j\rangle\langle\phi_j\vert$$
\begin{equation}
=\sum_{j=0}^N \delta_{2N}(2\phi_j-n_3)\vert\phi_j\rangle\langle\phi_j\vert
\end{equation}
Hence we have that $\hat{A}_L$ is a projection operator over a subspace generated by a subset of eigenvectors of the translation operator $\hat{T}(n_1,n_2)$.

\qee

\begin{examp}\label{ex_t_fourier_etc0}
For a line $L_q$ defined by $q=n_3$ (that is, $n_1=1$, $n_2=0$), the Wigner function summed over all point of $L_q$ is
\begin{equation}\label{estr_1}
\sum_{(q,p)\in L_q} W_{\rho}(q,p)=\sum_p W_{\rho}(n_3,p)=\langle n_3/2\vert\rho\vert n_3/2\rangle
\end{equation}
if $n_3$ is even, and equal to zero otherwise.
\end{examp}

\qee

More precisely, we have the following proposition:

\begin{pro}\label{lema_falta_paz}

Let $N$ be even and let $\rho$ be a density operator. Then
$$\sum_{p=0}^{2N-1} W_{\rho}(2q,p)= \langle q\vert\rho\vert q\rangle, \sp q=0,2,\dots,N-1$$
and
$$\sum_{p=0}^{2N-1} W_{\rho}(2q+1,p)= 0, \sp q=0,2,\dots,N-1$$
\end{pro}
{\bf Proof} First, to see why the case $q$ odd
implies that the Wigner function equals zero, consider the expression for $W$ given by
\begin{equation}\label{wewe1}
W_{\rho}(q,p)=\frac{1}{2N}\sum_{n=0}^{N-1}\langle q-n\vert\rho\vert n\rangle\exp\Big[\frac{2\pi i}{N}p(n-q/2) \Big]
\end{equation}
Write $\rho=\sum_j c_j\vert j\rangle\langle j\vert$, $c_j> 0$. Then
\begin{equation}
\langle q-n\vert\rho\vert n\rangle=\sum_j c_j\langle q-n\vert
j\rangle\langle j\vert n\rangle
\end{equation}
which is $\neq 0$ if and only if
$j=q-n=n$ for some $j$. In particular, in order to have a nonzero inner product above, we must have that $q$ is even, because $q-n=n$ implies $q=2n$.

\bigskip

Now suppose that $q=2q_0$. By the analysis above, we see that in the sum of the terms forming the Wigner function (eq. (\ref{wewe1})), we only have to sum the indices such that the equation
\begin{equation}\label{eq_modular1}
q-n=n \Leftrightarrow 2q_0-n=n
\end{equation}
is satisfied (recall that all calculations are made modulo N). Such equation has two solutions, namely $n=q_0$ and $n=q_0+N/2$. To see that there are no other solutions for (\ref{eq_modular1}), we proceed in the following way. From $2q_0-n=n$ we get $2(q_0-n)=0$. We know that $n=0$ and $n=q_0+N/2$ are solutions. Also, note that $x=0$ and $x=N/2$ are solutions of $2x=0$. Now, if  $y$ is a solution of $2x=0$ then $y-N/2$ also is. Clearly if $y$ is an element between $0$ and $N/2$ then $2y$ will be at most equal to $2N-2$, hence $2y\neq 0$. Finally, let $y$ be an element between $N/2$ and $N$ and by contradiction suppose that $2y=0$. Then by the remark above we have that $z=2y-N/2$ is also a solution and $z$ is between $0$ and $N/2$. But there are no solutions for $2x=0$ between $0$ and $N/2$. This shows that $2x=0$ admits only the solutions stated above.

\bigskip
Now note that if $n$ equals $q_0$ then
\begin{equation}
\exp\Big[\frac{2\pi i}{N}p(n-q/2) \Big]=1
\end{equation}
If $n=q_0+N/2$, we have that the exponential above is equal to $\pm 1$,
being positive or negative if $p$ is even or odd, respectively. Therefore, for $N$ even and $q=2q_0$, we have
\begin{equation}\label{paz_int_ex1}
W_{\rho}(2q_0,p)=\frac{1}{2N}(\langle q_0\vert \rho \vert
q_0\rangle\pm \langle q_0+N/2\vert \rho \vert q_0+N/2\rangle)
\end{equation}
where the sign $\pm$ depends on $p$. For fixed $q$ and considering all possible
$p$ (i.e., $p=0,\dots, 2N-1$), we have that the second inner product above will have a plus sign in front of it in the $N$ possibilities in which $p$ is even and will have a negative sign in the $N$ remaining possibilities. So
\begin{equation}
\sum_p W_{\rho}(2q_0,p)=\langle q_0\vert \rho \vert q_0\rangle
\end{equation}
This concludes the proof.
\qed

\begin{cor}\label{cor_apenas_pares}
If $q$ is odd then $W_{\rho}(q,p)=0$, for any $p$ and any $\rho$ density operator.
\end{cor}
{\bf Proof} Follows from the first paragraph of the proof above.

\qed

\begin{defi}
Let $\psi$ be a state. The {\bf $W$-transform} of $\psi$ is
\begin{equation}
\phi(p) :=\sum_{q=0}^{2N-1} W_{\psi}(q,2p)
\end{equation}
for $p=0,\dots,2N-1$.
\end{defi}
\bigskip

Let $\phi$ be the $W$-transform of $\psi$, and let
$\mathcal{F}\psi$ be the discrete Fourier transform of $\psi$.

\bigskip

{\bf Question}:
\begin{equation}
\vert(\mathcal{F}\psi)(p)\vert^2 \stackrel{?}{=} \phi(p), \sp p=0,1,\dots,N-1
\end{equation}
{\bf Answer}
For $N=2$ and $\psi=\vert 0\rangle$ or $\vert 1 \rangle$, the answer is yes. In fact, let $\vert\psi\rangle=\vert 0\rangle=(1,0)$. Then
$$\mathcal{F}\vert 0\rangle=\frac{1}{\sqrt{2}}\sum_j \exp{[2\pi ij0/2]}\vert j\rangle=\frac{1}{\sqrt{2}}(\vert 0\rangle +\vert 1\rangle)$$
$$\Rightarrow (\mathcal{F}\vert 0\rangle)(0)=\frac{1}{\sqrt{2}}\Rightarrow \vert(\mathcal{F}\vert 0\rangle)(0)\vert^2=\frac{1}{2}$$
And
$$\phi(0)=\sum_q W_{\vert 0\rangle}(q,0)=\frac{1}{4}+0+\frac{1}{4}+0=\frac{1}{2}$$
Also
$$(\mathcal{F}\vert 0\rangle)(1)=\frac{1}{\sqrt{2}}\Rightarrow \vert(\mathcal{F}\vert 0\rangle)(1)\vert^2=\frac{1}{2}$$
And
$$\phi(1)=\sum_q W_{\vert 0\rangle}(q,2)=\frac{1}{4}+0+\frac{1}{4}+0=\frac{1}{2}$$
Therefore in this case
\begin{equation}
\vert(\mathcal{F}\psi)(p)\vert^2 = \phi(p),\sp p=0,1
\end{equation}

\bigskip

Now let $\vert\psi\rangle=\vert 1\rangle=(0,1)$. Then
$$\mathcal{F}\vert 1\rangle=\frac{1}{\sqrt{2}}\sum_j \exp{[2\pi ij/2]}\vert j\rangle=\frac{1}{\sqrt{2}}(\vert 0\rangle +\exp{[2\pi i/2]}\vert 1\rangle)=\frac{1}{\sqrt{2}}(\vert 0\rangle -\vert 1\rangle)$$
$$\Rightarrow (\mathcal{F}\vert 1\rangle)(0)=\frac{1}{\sqrt{2}}\Rightarrow \vert(\mathcal{F}\vert 0\rangle)(0)\vert^2=\frac{1}{2}$$
And
$$\phi(0)=\sum_q W_{\vert 1\rangle}(q,0)=\frac{1}{4}+0+\frac{1}{4}+0=\frac{1}{2}$$
Also
$$(\mathcal{F}\vert 1\rangle)(1)=-\frac{1}{\sqrt{2}}\Rightarrow \vert(\mathcal{F}\vert 0\rangle)(0)\vert^2=\frac{1}{2}$$
And
$$\phi(1)=\sum_q W_{\vert 1\rangle}(q,2)=\frac{1}{4}+0+\frac{1}{4}+0=\frac{1}{2}$$
Therefore
\begin{equation}
\vert(\mathcal{F}\psi)(p)\vert^2 = \phi(p),\sp p=0,1
\end{equation}

Now let us write an example in which the state considered is mixed. Let
$\psi=1/\sqrt{2}(\vert 0\rangle+\vert 1\rangle)$. Then
\begin{equation}
\mathcal{F}\vert\psi\rangle=\frac{1}{\sqrt{2}}(\mathcal{F}\vert 0\rangle+\mathcal{F}\vert 1\rangle)=\frac{1}{\sqrt{2}}\Big[\frac{1}{\sqrt{2}}(\vert 0\rangle +\vert 1\rangle)+\frac{1}{\sqrt{2}}(\vert 0\rangle -\vert 1\rangle)\Big]=\vert 0\rangle
\end{equation}
Then $\vert(\mathcal{F}\psi)(0)\vert^2=1$ e $\vert(\mathcal{F}\psi)(1)\vert^2=0$. Now let us calculate $\phi(p)$, $p=0,1$. By definition, we have $\phi(p)=\sum_q W_{\psi}(q,2p)$. We can use the expression (\ref{fech_mist}):
\begin{equation}
W_{\psi}(q,0)=\frac{1}{2}\Big(W_{\vert 0\rangle}(q,0)+W_{\vert 1\rangle}(q,0)+\Delta_{0,1}(q,0)\Big)
\end{equation}
\begin{equation}
W_{\psi}(q,2)=\frac{1}{2}\Big(W_{\vert 0\rangle}(q,2)+W_{\vert 1\rangle}(q,2)+\Delta_{0,1}(q,2)\Big)
\end{equation}
Then
$$W_{\psi}(0,0)=\frac{1}{2}(\frac{1}{4}+\frac{1}{4}+0)=\frac{1}{4}$$
$$W_{\psi}(1,0)=\frac{1}{2}(0+0+\frac{1}{2})=\frac{1}{4}$$
$$W_{\psi}(2,0)=\frac{1}{2}(\frac{1}{4}+\frac{1}{4}+0)=\frac{1}{4}$$
$$W_{\psi}(3,0)=\frac{1}{2}(0+0+\frac{1}{2})=\frac{1}{4}$$
which implies $\phi(0)=1=\vert(\mathcal{F}\psi)(0)\vert^2$.
Similarly,
$$W_{\psi}(0,2)=\frac{1}{2}(\frac{1}{4}+\frac{1}{4}+0+0)=\frac{1}{4}$$
$$W_{\psi}(1,2)=\frac{1}{2}(0+0-\frac{1}{2})=-\frac{1}{4}$$
$$W_{\psi}(2,2)=\frac{1}{2}(\frac{1}{4}+\frac{1}{4}+0+0)=\frac{1}{4}$$
$$W_{\psi}(3,2)=\frac{1}{2}(0+0-\frac{1}{2})=-\frac{1}{4}$$
and so $\phi(1)=0=\vert(\mathcal{F}\psi)(1)\vert^2$.

\qee

Inspired in the calculation above, we prove the following lemma, valid for pure states only. After that, we will prove the result for density operators.

\bigskip

\begin{lem}
Let $\psi=\vert m\rangle\in\{\vert 0\rangle,\dots ,\vert N-1\rangle\}$, $N$ even. Then
\begin{equation}
\vert(\mathcal{F}\psi)(p)\vert^2 = \phi(p), \sp p=0,1,\dots,N-1
\end{equation}
\end{lem}
{\bf Proof} We have
$$\mathcal{F}\vert m\rangle=\frac{1}{\sqrt{N}}\sum_{j=0}^{N-1}\exp{[2\pi ijm/N]}\vert j\rangle$$
So
$$(\mathcal{F}\vert m\rangle)(p)=\frac{1}{\sqrt{N}}\exp{[2\pi ipm/N]}\Rightarrow \vert(\mathcal{F}\vert m\rangle)(p)\vert^2=\frac{1}{N}$$
Let us calculate $\phi(p)=\sum_{q=0}^{2N-1} W_{\vert
m\rangle}(q,2p)$. By the corollary \ref{cor_apenas_pares}, we only
have to sum the even $q$. Then $\phi(p)=\sum_{q=0}^{N-1} W_{\vert
m\rangle}(2q,2p)$. By proposition \ref{lema_falta_paz} we get,
using expression (\ref{paz_int_ex1}), that
\begin{equation}
W_{\rho}(2q_0,p)=\frac{1}{2N}(\langle q_0\vert \rho \vert
q_0\rangle+ \langle q_0+N/2\vert \rho \vert q_0+N/2\rangle)
\end{equation}
where the sign of the second inner product is positive because $2p$ is even. Now note that only one of the inner products above can be nonzero, because $\rho$ is pure, by assumption. Moreover, $\rho$ pure implies that such inner products are equal to 1. Finally, since $q$ varies between $0$ and $2N-1$ we have exactly two nonzero terms in the sum of $\phi(p)$ namely, the terms corresponding to the $m$ and $m+N/2$ indices. Hence,
$$\phi(p)=1/2N+1/2N=1/N=\vert(\mathcal{F}\vert m\rangle)(p)\vert^2$$
This concludes the proof.

\qed

The following result, inspired in the previous one, completes proposition \ref{lema_falta_paz}, which related the discrete Wigner function with the base of position vectors. Now we do the corresponding work for the basis of momentum vectors.

\begin{pro}\label{pro_paz2} Let $N$ be even and let $\rho$ be a density operator. Let $\vert p\rangle$ be a vector of the momentum basis, that is, obtained via the discrete Fourier transform of a position base vector:
\begin{equation}
\vert p\rangle=\frac{1}{\sqrt{N}}\sum_{j=0}^{N-1}\exp{[2\pi ijp/N]}\vert j\rangle
\end{equation}
Then
\begin{equation}
\sum_{q=0}^{2N-1} W_{\rho}(q,2p)=\langle p\vert \rho\vert p\rangle, \sp p=0,1,\dots N-1
\end{equation}
\begin{equation}
\sum_{q=0}^{2N-1} W_{\rho}(q,2p+1)=0, \sp p=0,1,\dots N-1
\end{equation}
\end{pro}
{\bf Proof} Let us calculate $\phi(p)=\sum_{q=0}^{2N-1} W_{\rho}(q,2p)$. By corollary \ref{cor_apenas_pares}, we only have to sum the even $q$ indices. Then $\phi(p)=\sum_{q=0}^{N-1} W_{\rho}(2q,2p)$. By proposition \ref{lema_falta_paz} we get, using expression (\ref{paz_int_ex1}), that
\begin{equation}
W_{\rho}(2q,2p)=\frac{1}{2N}(\langle q\vert \rho \vert
q\rangle+ \langle q+N/2\vert \rho \vert q+N/2\rangle)
\end{equation}
where the sign of the second inner product is a plus because $2p$ is even. Write $\rho=\sum_i c_i\vert i\rangle\langle i\vert$. Take, for instance, $q=0$. Then
$$W_{\rho}(0,2p)=\frac{1}{2N}(\langle 0\vert \rho \vert
0\rangle+ \langle 0+N/2\vert \rho \vert 0+N/2\rangle)$$
\begin{equation}
=\frac{1}{2N}(\sum_{i} c_i\langle 0\vert i\rangle\langle i \vert
0\rangle+ \langle N/2\vert i\rangle\langle i \vert N/2\rangle)=\frac{1}{2N}(c_0+c_{N/2})
\end{equation}
As we know, $W_{\rho}(1,2p)=0$. Take $q=2$, then
\begin{equation}
W_{\rho}(2,2p)=\frac{1}{2N}(c_1+c_{N/2+1})
\end{equation}
and so on (noting that we always have zeroes when $q$ is odd). In this way, we end up summing all $c_i$ coefficients $c_i$ twice  (because $q$ varies between 0 and $2N-1$) and we get that
\begin{equation}
\phi(p)=\sum_{q=0}^{2N-1} W_{\rho}(q,2p)=\frac{1}{N}(c_0+c_1+\cdots +c_{2N-1})=\frac{1}{N}
\end{equation}
By the calculation above, we only have to calculate $\langle p\vert \rho\vert p\rangle$ and show that such number equals $1/N$. Recall that the inner product we consider is linear on the second variable, so we write $\rho=\sum_m c_m\vert m\rangle\langle m\vert$ and then:
$$\langle p\vert \rho\vert p\rangle=\sum_m c_m \frac{1}{N}\sum_{j=0}^{N-1} \exp{[-2\pi ijp/N]}\sum_{l=0}^{N-1}\exp{[2\pi ilp/N]}\langle j\vert m\rangle\langle m\vert l\rangle$$
\begin{equation}
=\sum_m c_m \frac{1}{N}\sum_{j=0}^{N-1} \exp{[-2\pi ijp/N]}\exp{[2\pi imp/N]}\langle j\vert m\rangle=\frac{1}{N}\sum_m c_m=\frac{1}{N}
\end{equation}
\qed {\bf Conclusion} By propositions \ref{lema_falta_paz} and
\ref{pro_paz2} we have for the discrete Wigner transform that if
$N$ is even and $\rho$ is a density operator then
\begin{equation}
\sum_{p=0}^{2N-1} W_{\rho}(2q,p)= \langle q\vert\rho\vert q\rangle, \sp
\sum_{p=0}^{2N-1} W_{\rho}(2q+1,p)= 0, \sp q=0,1,\dots,N-1
\end{equation}
and if
\begin{equation}
\vert p\rangle=\frac{1}{\sqrt{N}}\sum_{j=0}^{N-1}\exp{[2\pi ijp/N]}\vert j\rangle
\end{equation}
then
\begin{equation}
\sum_{q=0}^{2N-1} W_{\rho}(q,2p)=\langle p\vert \rho\vert p\rangle, \sp
\sum_{q=0}^{2N-1} W_{\rho}(q,2p+1)=0, \sp p=0,1,\dots N-1
\end{equation}
Such expressions are the discrete analog of the result we have for the continuous Wigner function, namely the result that relates the marginals with the Fourier transform $\mathcal{F}$: if $\rho=\vert \psi\rangle\langle\psi\vert$ then
\begin{equation}
\int W_\rho(q,p) dp=\vert\psi(q)\vert^2,\sp \int W_\rho(q,p) dq=\vert\mathcal{F}\psi(p)\vert^2
\end{equation}
See \cite{gosson} for more details.

\qee

\begin{examp}
Denote by $\mathcal{W}_\rho$ the matrix with entries $W_\rho(q,p)$ for $q, p=0,\dots, 2N-1$. For instance, if $N=2$ and writing $\vert 0 \rangle=(1,0)$ and $\vert 1 \rangle=(0,1)$, we have that $\mathcal{W}_{\rho}$ contains the image of the Wigner function for each point of the phase space. We immediately notice that the integral over all space equals 1:
\begin{equation}
\mathcal{W}_{\vert 0\rangle\langle 0\vert}=\left(
\begin{array}{cccc}
\frac{1}{4} & \frac{1}{4} & \frac{1}{4} & \frac{1}{4} \\
0 & 0 & 0 & 0 \\
\frac{1}{4} & -\frac{1}{4} & \frac{1}{4} & -\frac{1}{4} \\
0 & 0 & 0 & 0
\end{array}
\right), \sp \mathcal{W}_{\vert 1\rangle\langle 1\vert}=\left(
\begin{array}{cccc}
\frac{1}{4} & -\frac{1}{4} & \frac{1}{4} & -\frac{1}{4} \\
0 & 0 & 0 & 0 \\
\frac{1}{4} & \frac{1}{4} & \frac{1}{4} & \frac{1}{4} \\
0 & 0 & 0 & 0
\end{array}
\right)
\end{equation}
\end{examp}
\qee

\begin{examp}
Denote by $\mathcal{W}_\rho$ the matrix with entries $W_\rho(q,p)$ for $q, p=0,\dots, 2N-1$. Let $N=4$, and writing $\vert 0 \rangle=(1,0,0,0)$, $\vert 1 \rangle=(0,1,0,0)$, $\vert 2 \rangle=(0,0,1,0)$, $\vert 3 \rangle=(0,0,0,1)$, we have, in a similar way as seen in the previous example, that
$$\mathcal{W}_{\vert 0\rangle\langle 0\vert}=\left(
\begin{array}{cccccccc}
\frac{1}{8} & \frac{1}{8} & \frac{1}{8} & \frac{1}{8} & \frac{1}{8} & \frac{1}{8} & \frac{1}{8} & \frac{1}{8} \\
0 & 0 & 0 & 0 & 0 & 0 & 0 & 0 \\
0 & 0 & 0 & 0 & 0 & 0 & 0 & 0 \\
0 & 0 & 0 & 0 & 0 & 0 & 0 & 0 \\
\frac{1}{8} & -\frac{1}{8} & \frac{1}{8} & -\frac{1}{8} & \frac{1}{8} & -\frac{1}{8} & \frac{1}{8} & -\frac{1}{8} \\
0 & 0 & 0 & 0 & 0 & 0 & 0 & 0 \\
0 & 0 & 0 & 0 & 0 & 0 & 0 & 0 \\
0 & 0 & 0 & 0 & 0 & 0 & 0 & 0 \\
\end{array}
\right)$$
$$\mathcal{W}_{\vert 1\rangle\langle 1\vert}=\left(
\begin{array}{cccccccc}
0 & 0 & 0 & 0 & 0 & 0 & 0 & 0 \\
0 & 0 & 0 & 0 & 0 & 0 & 0 & 0 \\
\frac{1}{8} & \frac{1}{8} & \frac{1}{8} & \frac{1}{8} & \frac{1}{8} & \frac{1}{8} & \frac{1}{8} & \frac{1}{8} \\
0 & 0 & 0 & 0 & 0 & 0 & 0 & 0 \\
0 & 0 & 0 & 0 & 0 & 0 & 0 & 0 \\
0 & 0 & 0 & 0 & 0 & 0 & 0 & 0 \\
\frac{1}{8} & -\frac{1}{8} & \frac{1}{8} & -\frac{1}{8} & \frac{1}{8} & -\frac{1}{8} & \frac{1}{8} & -\frac{1}{8} \\
0 & 0 & 0 & 0 & 0 & 0 & 0 & 0 \\
\end{array}
\right)$$
$$\mathcal{W}_{\vert 2\rangle\langle 2\vert}=\left(
\begin{array}{cccccccc}
\frac{1}{8} & -\frac{1}{8} & \frac{1}{8} & -\frac{1}{8} & \frac{1}{8} & -\frac{1}{8} & \frac{1}{8} & -\frac{1}{8} \\
0 & 0 & 0 & 0 & 0 & 0 & 0 & 0 \\
0 & 0 & 0 & 0 & 0 & 0 & 0 & 0 \\
0 & 0 & 0 & 0 & 0 & 0 & 0 & 0 \\
\frac{1}{8} & \frac{1}{8} & \frac{1}{8} & \frac{1}{8} & \frac{1}{8} & \frac{1}{8} & \frac{1}{8} & \frac{1}{8} \\
0 & 0 & 0 & 0 & 0 & 0 & 0 & 0 \\
0 & 0 & 0 & 0 & 0 & 0 & 0 & 0 \\
0 & 0 & 0 & 0 & 0 & 0 & 0 & 0 \\
\end{array}
\right)$$
$$\mathcal{W}_{\vert 3\rangle\langle 3\vert}=\left(
\begin{array}{cccccccc}
0 & 0 & 0 & 0 & 0 & 0 & 0 & 0 \\
0 & 0 & 0 & 0 & 0 & 0 & 0 & 0 \\
\frac{1}{8} & -\frac{1}{8} & \frac{1}{8} & -\frac{1}{8} & \frac{1}{8} & -\frac{1}{8} & \frac{1}{8} & -\frac{1}{8} \\
0 & 0 & 0 & 0 & 0 & 0 & 0 & 0 \\
0 & 0 & 0 & 0 & 0 & 0 & 0 & 0 \\
0 & 0 & 0 & 0 & 0 & 0 & 0 & 0 \\
\frac{1}{8} & \frac{1}{8} & \frac{1}{8} & \frac{1}{8} & \frac{1}{8} & \frac{1}{8} & \frac{1}{8} & \frac{1}{8} \\
0 & 0 & 0 & 0 & 0 & 0 & 0 & 0 \\
\end{array}
\right)$$

\end{examp}
\qee

{\bf Remark 1} What occurs in general for pure states: the Wigner function $W_{\vert q_0\rangle\langle q_0\vert}$ is zero except in two lines, located in $q \equiv 2(mod N)$. When $q=2q_0$, $W$ assumes the value $1/2N$, and when $q=2q_0\pm N$, $W$ assumes the value $1/2N$ for even values of $p$ and $-1/2N$ for odd values. Such oscillations are typical of interference fringes and can be interpreted as arising from the interference between the line $q=2q_0$ and a mirror image formed at a distance of $2N$ from $2q_0$, induced by the periodic boundary conditions \cite{miquel}.

\bigskip

{\bf Remark 2} The fact that the Wigner function assumes negative values in the interference line is essential for one to be able to recover the correct marginal distributions. Summing the values $W(q,p)$ along a vertical line gives us the probability  of measuring $q/2$, which should be equal to 1 if $q=2q_0$, and equal to zero, otherwise.

\qee

A natural question is to try to understand the action of the
operator which defines QIFS in the dual variables $p$. This is the
purpose of the next results.

\begin{lem}\label{um_lema_dcomut1}
Let $\Lambda(\rho)=\sum_i V_i\rho V_i^*$ and define $F(\rho)=\mathcal{F}\rho\mathcal{F}^*$, where $\mathcal{F}$ is any unitary map. Then there is $G:\mathcal{M}_N\to\mathcal{M}_N$ such that the above diagram commutes:
\begin{equation}
\begin{CD}
\mathcal{M}_N @>F>> \mathcal{M}_N\\
@V\Lambda VV @VVGV \\
\mathcal{M}_N @>F>> \mathcal{M}_N
\end{CD}
\end{equation}
\end{lem}
{\bf Proof} First, note that $F^{-1}(\rho)=\mathcal{F}^{*}\rho\mathcal{F}$. Also $\mathcal{F}$ is unitary, therefore we have $\mathcal{F}^{-1}=\mathcal{F}^{*}$. Define $G=F\circ\Lambda\circ F^{-1}$. Explicitly,
$$G(\rho)=F(\sum_i V_i\mathcal{F}^*\rho \mathcal{F}V_i^*)=\mathcal{F}\Big[ \sum_i V_i\mathcal{F}^*\rho \mathcal{F}V_i^*\Big] \mathcal{F}^*$$
$$=\sum_i \mathcal{F}V_i\mathcal{F}^*\rho \mathcal{F}V_i^*\mathcal{F}^* =\sum_i \tilde{V}_i\rho \tilde{V}_i^*$$
where $\tilde{V}_i=\mathcal{F}V_i\mathcal{F}^*$. And a simple inspection shows that
$$F(\Lambda(\rho))=G(F(\rho))=\sum_i \mathcal{F}V_i\rho V_i^*\mathcal{F}^*$$

\qed

\begin{exe} Consider $N=2$. Then the discrete Fourier transform is given by
\begin{equation}
\mathcal{F}=\frac{1}{\sqrt{2}}\left(
\begin{array}{cc}
1 & 1 \\
1 & -1
\end{array}
\right)
\end{equation}
In this case we have $\mathcal{F}^{-1}=\mathcal{F}$. Let
\begin{equation}
V_1=\left(
\begin{array}{cc}
\sqrt{p_{11}} & 0\\
0 & 0
\end{array}
\right),
\sp V_2=\left(
\begin{array}{cc}
0 & \sqrt{p_{12}}\\
0 & 0
\end{array}
\right),
\end{equation}
\begin{equation}
 V_3=\left(
\begin{array}{cc}
\sqrt{p_{21}} & 0\\
0 & 0
\end{array}
\right),
\sp V_4=\left(
\begin{array}{cc}
0 & 0\\
0 & \sqrt{p_{22}}
\end{array}
\right)
\end{equation}
where  the $p_{ij}$ form a column stochastic matrix $P$.  Then lemma \ref{um_lema_dcomut1} for this example shows that $G(\rho)=\sum_i \tilde{V}_i\rho \tilde{V}_i^*$, where
$$
\tilde{V}_1=\mathcal{F}V_1\mathcal{F}^*=\frac{1}{2}\sqrt{p_{11}}\left(
\begin{array}{cc}
1 & 1\\
1 & 1
\end{array}
\right),\sp\tilde{V}_2=\mathcal{F}V_2\mathcal{F}^*=\frac{1}{2}\sqrt{p_{12}}\left(
\begin{array}{cc}
1 & -1\\
1 & -1
\end{array}
\right)$$
$$\tilde{V}_3=\mathcal{F}V_3\mathcal{F}^*=\frac{1}{2}\sqrt{p_{21}}\left(
\begin{array}{cc}
1 & 1\\
-1 & -1
\end{array}
\right),\sp\tilde{V}_4=\mathcal{F}V_4\mathcal{F}^*=\frac{1}{2}\sqrt{p_{22}}\left(
\begin{array}{cc}
1 & -1\\
-1 & 1
\end{array}
\right)
$$
Then, from $p_{11}+p_{21}=1$, $p_{12}+p_{22}=1$ and writing
$$\rho=\left(
\begin{array}{cc}
\rho_{11} & \rho_{12}\\
\rho_{21} & 1-\rho_{11}
\end{array}
\right)
$$
we get from \ref{um_lema_dcomut1} the expression
$$F(\Lambda(\rho))=G(F(\rho))=\sum_i \mathcal{F}V_i\rho V_i^*\mathcal{F}^*$$
\begin{equation}
=\left(
\begin{array}{cc}
\frac{1}{2} & p_{11}\rho_{11}+p_{12}(1-\rho_{11})-\frac{1}{2}\\
p_{11}\rho_{11}+p_{12}(1-\rho_{11})-\frac{1}{2} & \frac{1}{2}
\end{array}
\right)
\end{equation}
In the case that the vector $\pi=(\rho_{11},1-\rho_{11})$ is fixed for the stochastic matrix $P$, we can rewrite the expression above as
\begin{equation}
F(\Lambda(\rho))=G(F(\rho))
=\left(
\begin{array}{cc}
\frac{1}{2} & \rho_{11}-\frac{1}{2}\\
\rho_{11}-\frac{1}{2} & \frac{1}{2}
\end{array}
\right)
\end{equation}

\end{exe}

\qee

\begin{lem}
Define $\Lambda:\mathcal{M}_N\to \mathcal{M}_N$, $\Lambda(\rho)=\sum_{i} V_i\rho V_i^*$, with $V_i$ linear, $\sum_i V_i^*V_i=I$ and let
$W_{\Lambda(\rho)}$ be the associated discrete Wigner function.
Then given $(q,p)$ there are $M_i=M_i(q,p)$ such that
$$W_{\Lambda(\rho)}(q,p)=\sum_i tr(M_i\rho M_i^*)$$
\end{lem}
{\bf Proof} First, as $A(q,p)$ is hermitian, we have a decomposition
$$A=UDU^{-1}$$
where $U$ is unitary and $D$ is diagonal (and real). Then
$$A^{1/2}=UD^{1/2}U^{-1}$$
where $(A^{1/2})^2=A$, $D^{1/2}$ is the diagonal matrix whose entries are the positive square roots of the entries of $D$. Then
$$
W_{\Lambda(\rho)}(q,p)=tr(\hat{A}(q,p)\Lambda(\rho))=tr(\hat{A}\sum_i V_i\rho V_i^*)=\sum_i tr(\hat{A} V_i\rho V_i^*)$$
\begin{equation}
=\sum_i tr(A^{1/2} V_i\rho V_i^*A^{1/2})=\sum_i tr(UD^{1/2}U^{-1} V_i\rho V_i^*UD^{1/2}U^{-1})
\end{equation}
Defining $M_i=UD^{1/2}U^{-1} V_i$ and noting that $U^{-1}=U^*$, we can write
$$W_{\Lambda(\rho)}(q,p)=\sum_i tr(M_i\rho M_i^*)$$
\qed

\end{document}